# Zero-Magnetic Field Fractional Quantum States

S. Kumar[1,2,*], M. Pepper[1,2], S. N. Holmes[3], H. Montagu[1,2], Y. Gul[1,2], D. A. Ritchie[4], I. Farrer[4,5]

[1]*London Centre for Nanotechnology, 17-19 Gordon Street, London WC1H 0AH, United Kingdom*

[2]*Department of Electronic and Electrical Engineering, University College London, Torrington Place, London WC1E 7JE, United Kingdom*

[3]*Toshiba Research Europe Limited, Cambridge Research Laboratory, 208 Cambridge Science Park, Milton Road, CB4 0GZ, United Kingdom*

[4]*Cavendish Laboratory, J.J. Thomson Avenue, Cambridge CB3 OHE, United Kingdom*

5*Now at Department of Electronic and Electrical Engineering, University of Sheffield, Mappin Street, Sheffield S1 3JD, United Kingdom*

**Abstract**

Since the discovery of the Fractional Quantum Hall Effect in 1982 there has been considerable theoretical discussion on the possibility of fractional quantization of conductance in the absence of Landau levels formed by a quantizing magnetic field. Although various situations have been theoretically envisaged, particularly lattice models in which band flattening resembles Landau levels, the predicted fractions have never been observed. In this Letter, we show that odd and even denominator fractions can be observed, and manipulated, in the absence of a quantizing magnetic field, when a low-density electron system in a GaAs based one-dimensional quantum wire is allowed to relax in the second dimension. It is suggested that such a relaxation results in formation of a zig-zag array of electrons with ring paths which establish a cyclic current and a resultant lowering of energy. The behavior has been observed for both symmetric and asymmetric confinement but increasing the asymmetry of the confinement potential, to result in a flattening of confinement, enhances the appearance of new fractional states. We find that an in-plane magnetic field induces new even denominator fractions possibly indicative of electron pairing. The new quantum states described here have implications both for the physics of low dimensional electron systems and also for quantum technologies. This work will enable further development of structures which are designed to electrostatically manipulate the electrons for the formation of particular configurations. In turn, this could result in a designer tailoring of fractional states to amplify particular properties of importance in future quantum computation.

The discovery of the Integer and Fractional Quantum Hall Effects, IQHE and FQHE, in two-dimensional (2D) systems in the presence of a quantizing magnetic field stimulated an enormous theoretical and experimental activity [1-6]. The first suggestion that a magnetic field was not essential was from Haldane [7] who considered a filled, tight binding, band based on a honeycomb lattice. This system would show the quantization if time reversal symmetry were to be broken by a spatially inhomogeneous magnetic field whose average value was zero. Numerical calculations showed that a band formed from a regular tight binding lattice could resemble a Landau level by being sufficiently flat, and almost dispersionless, so that the kinetic energy is minimized. In this situation, it was found that a localized wavefunction model predicted energy minima at filling factors of 1/3 and 1/5 on the basis of nearest and next nearest neighbor interactions [8]. In this respect, such lattice models could resemble a Composite Fermion model [9], in which flux quanta act as lattice sites. However, there were no reports of experimental observations of fractional quantization of conductance until very recently when quasi-1D hole conduction in Germanium quantum wires [10], corresponded to charge values of e/2 and e/4.

A 1D system, defined by means of split gate [11] with an additional top gate on a 2D electron gas, exhibits quantized conductance at values of $2ne^2/h$, where n is an integer 1,2,3,…[12,13]. If the confinement is reduced, so allowing the electrons to relax in the second dimension, the energy levels are determined by both the electron-electron interaction and the spatial confinement. This situation has been investigated in detail both experimentally [14-16] and theoretically [17-19]. It is found that as the role of the interaction (confinement) increases (decreases), so a line of electrons adopts a zig-zag or two-row configuration to minimize their mutual repulsion [17], which has been imaged recently [20]. In the present work, we have utilized quasi-1D electrons and show that there is a range of parameter space such as confinement conditions and carrier concentration which support a mix of conductance fractionalization in the relaxed 1D configuration [21].

The devices used in the present work were fabricated from GaAs/AlGaAs heterostructures grown using molecular beam epitaxy. The two-dimensional electron gas (2DEG) formed up to 290 nm beneath the surface had a low-temperature mobility of $1.6 \times 10^6$ cm$^2$/V.s ($2.0 \times 10^6$ cm$^2$/V.s), and an electron density of $9.0 \times 10^{10}$ cm$^{-2}$ ($1.0 \times 10^{11}$ cm$^{-2}$) for sample 1 (sample 2) [Supplementary Material, Ref. 22]. A pair of split gates of length (width) of 0.4 µm (0.7 µm), and a top gate of length 1 µm separated by a 300 nm thick insulating layer of cross-linked poly(methyl methacrylate) (PMMA) were patterned by a standard lithographic technique, see the cartoon in the inset, Fig. 1(a) [16]. In all experiments, the two-terminal differential conductance ($G$) measurements in the Ohmic and non-Ohmic (non-linear) regimes were performed by sweeping the split gate voltage, $V_{sg}$ in the presence of an excitation voltage of 10 µV at 73 Hz in a cryofree dilution refrigerator with a lattice temperature of 25 mK [23,24].

Figure 1(a) shows results on a 1D channel where $G$ is measured as a function of increasing the asymmetry of the confinement potential by applying an offset, $\Delta V_{sg}$, to one of the split gates with the two gates then being swept together. The net effect of the offset was to push the conducting channel sideways and increase its width. The top gate voltage, $V_{tg}$ is kept constant at -0.36 V, the initial 2D carrier concentration, $n_{2D}$ was $5 \times 10^{10}$ cm$^{-2}$. In the discussion of this and subsequent figures, all fractions are in units of $e^2/h$, and the fractional states corresponding to the flattest plateaux are labelled in the right vertical axis. On the left, the 0.7($2e^2/h$) structure, appearing at 1.5($e^2/h$), and a change of gradient at 1 are present. The 0.7($2e^2/h$) structure disappears with an increase in channel width when a structure emerges near 2 which gradually drops to flatten very close to 3/5, as confinement is further weakened it persists in the vicinity before disappearing. The structure at 1 (extreme left) rapidly drops on increasing the channel width to form a distinct plateau at 1/6 where it remains with increasing width before turning into a structure which increases in value until merging with the plateau at 3/5, this then drops with increasing width to become flat at 2/5, shown in green trace. The flattening at 2/5 was a minimum in the conductance as the structure then increased with increasing width with a minimum gradient at 1/2 before disappearing at which stage the conduction rises rapidly indicating that a 2D transition has occurred.

We define a plateau by a minimum in gradient (increasing flatness) whereas the change between plateau values has a distinct slope. Although we consider that stable quantum states are indicated by the flattest plateaux there are cases in this Letter where a plateau starts with a slope and then contains a flat part. Occasionally there is a slope which remains constant with a change in gate voltages, possibly this may be an indication of the stability of the fractional states. However, here we just consider the minimum gradient structure.

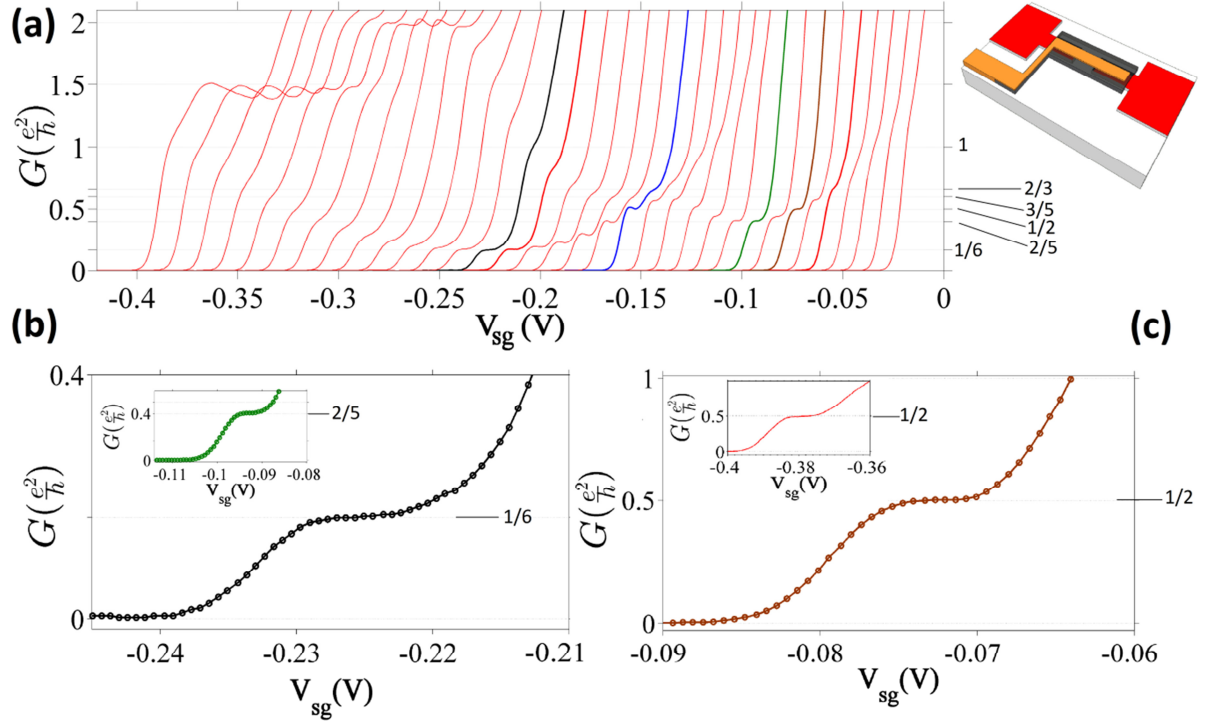

*Figure 1: Conductance characteristics of a top-gated split gate device as a function of asymmetry and width of the confinement potential (sample-1). Inset on the right side in (a) shows a schematic diagram of the device; split gates are shown in red; PMMA is shown in grey sandwiched between the split gates and the top gate (orange). (a) At $V_{tg}$=-0.36 V, the offset or asymmetry between the split gates, $\Delta V_{sg}$ was incremented from 0.65 V (left) to 0.98 V (right) in steps of 10 mV so widening the channel and flattening the confinement. (b) main and inset, and (c) show close up views of fractional states at 1/6, 2/5 and 1/2, respectively as seen in (a). An inset in (c) shows a close up of the 1/2 fractional state in a symmetric confinement potential in sample 2. A horizontal offset of 4 mV was set between the consecutive traces in (a) for clarity.*

We consistently found the following as stable states 1/6, 2/5, 1/2; a 2/3 plateau was frequently found with a slight slope, and the 3/5 plateau was often present but slightly below the expected fractional value. We estimated the accuracy of plateaux corresponding to the flattest fractions was 1 in $10^4$ for a split gate voltage length for the 1/6, 1/2, and 2/5 fractional states, respectively. The absolute value and length of plateau for these fractions is over 2 mV, (error 0.8%), 3.3 mV, (error 0.2 %), and 3.4 mV, (error 0.2 %), respectively. It is noteworthy that whereas the odd denominator fractions have been observed in the FQHE, the 1/6 and 1/2 have not been observed before although a 1/2 is found in a higher Landau level, 5/2. The main plot and inset in Fig. 1(b), and Fig. 1(c) show a close-up of the 1/6, 2/5 and 1/2 plateaux in an asymmetric confinement potential [from Fig. 1(a)]. Also, we observed the fractional plateau at 1/2 for asymmetric confinement potential in sample 2 [inset, Fig. 1(c)].

Figure 2 illustrates the sensitivity of the system to a change in carrier concentration at a fixed in-plane magnetic field, $B_{II}$=10 T normal to the current direction. A fixed offset on the split gate voltage produces an asymmetric, wide confinement and the higher integer plateaux 1,2,3($e^2$/h) are not found. Each plot corresponds to a more negative top gate voltage producing a decreasing $n_{2D}$ from left to right in the range (4.0-1.0)x10$^{10}$ cm$^{-2}$ [See Supplementary Material, Ref. 22]. It is seen that a weak structure on the left, drops in value until it flattens on either side of a conductance value of 1/2. The traces on either side of the indexed 1/2 exhibit flat plateau in the range 6/11 to 5/11, we suggest that this indicates a band of stable states within these limits. Outside the limits, the structure develops a gradient which alternates in sign. On further reducing the carrier concentration, the structure drops in value and flattens before disappearing indicating a transition to 2D behavior for maximum channel width. Flat plateaux form on both sides of the 1/4 between limits of 3/10 and 6/25 indicating that there is a band of stable states which are located around the 1/4 in a similar way to the states in the vicinity of 1/2. There is a resemblance here to the observations reported for holes in Ge [10]. The theoretical discussion on the 5/2 plateau in the FQHE suggests its origin to stem from a pairing of two 1/4 states [25] and has stimulated considerable interest in Quantum Technology applications as it is thought to be Non-Abelian. The similarity in the behavior of the structure near the 1/2 and 1/4 values found here indicates that they are linked. The band of states around the 1/4 and 1/2 may be related to the proposed non-Abelian nature of the 1/4 in that the differing details of the exchange path provide distinct states which are degenerate. The differences due to the small changes in the confinement provide a different exchange path, the change in confinement energy alters the level energy to provide the band around the 1/4. Similar considerations apply to the formation of the band around the 1/2, particularly if this arises from pairs of 1/4 states.

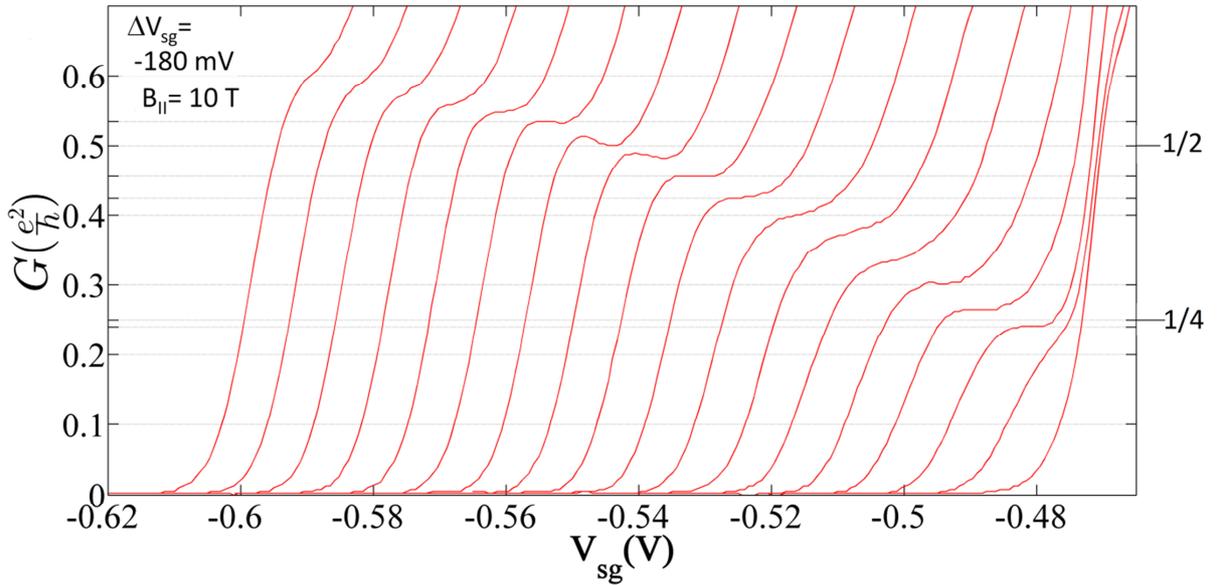

*Figure 2: Effect of variation of carrier concentration on the conductance in the presence of $B_{II}$ (sample-2). $V_{tg}$ was varied from -0.45 V (left) to -0.62 V (right) in steps of -10 mV, at $B_{II}$=10 T; $\Delta V_{sg}$ =-180 mV.*

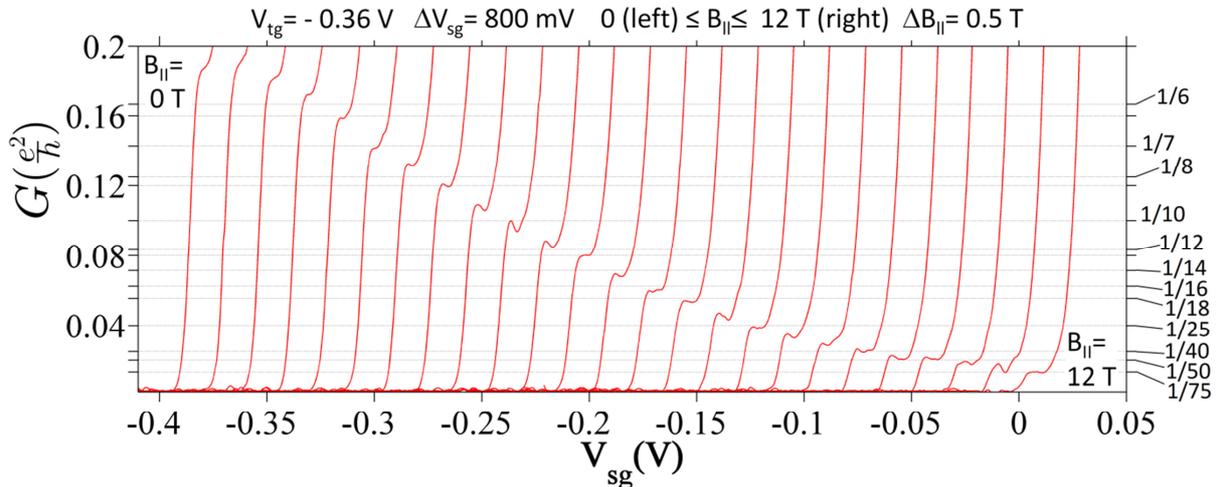

*Figure 3: Effect of variation of $B_{II}$ on the fractional states (Sample-1). $B_{ll}$ was varied from 0 T (left) to 12 T (right) in steps of 0.5 T; $V_{tg}$= -0.36 V; $\Delta V_{sg}$= 800 mV.*

The sensitivity to the magnitude of in-plane magnetic field is shown in more detail in Fig. 3, where both $n_{2D}$=5x10$^{10}$ cm$^{-2}$ and the offset voltage are held constant. The structure on the left commences movement when the field is applied and continues to decrease as the field is increased. The gradient of the structure varies continually with increasing $B_{||}$ but there are regions of flatness, or minimal gradient, which occur in the vicinity of 1/6, 1/12, 1/18 and, surprisingly, near 1/25, at the highest values of B the conductance drops sharply and much smaller values of conductance structure are observed. The continual movement of the structure with $B_{||}$, from a small value, implies that it is not related to spin but, rather, is dependent on flux.

In order to investigate the nature of the conduction process, we have measured the differential conductance as the source-drain voltage ($V_{sd}$) was increased into the non-Ohmic regime. It is a well-known fact that the differential conductance becomes one-half of its value in the non-Ohmic regime due to the lifting of the momentum degeneracy [26-27]. Figure 4(a) shows a conductance plateau at 2/5 at $V_{sd}$=0 V(right); on increasing $V_{sd}$ the structure eventually flattens and settles at 1/5. This result indicates a fractional band conduction is occurring as opposed to a fractional transmission coefficient which will behave in a different manner with source-drain voltage and tend to increase so increasing the differential conductance. Also, as the differential conductance decreased towards the 1/5 a plateau emerged at 1/4 perhaps indicating that another state was attempting to form as the conducting level was pulled further down below the source potential and the carrier concentration increased. However, the observation of the 1/5 indicates the stability of this state for the confinement conditions.

Figure 4(b) shows the differential conductance of an Ohmic fractional plateau at 1/2. The differential conductance drops, as expected, with increasing $V_{sd}$ but attempts to flatten and form a plateau at 1/3 followed by a flattening at 3/10 with intervening structure. The structure always remains slightly above the value of 1/4 produced by the lifting of the momentum degeneracy. The conversion of the 1/2 into a non-Ohmic differential of 1/3 implies that this state is derived from an Ohmic 2/3; the states then appear as if almost degenerate with only small changes in energy required for a conversion process [See Supplementary Material, Ref. 22].

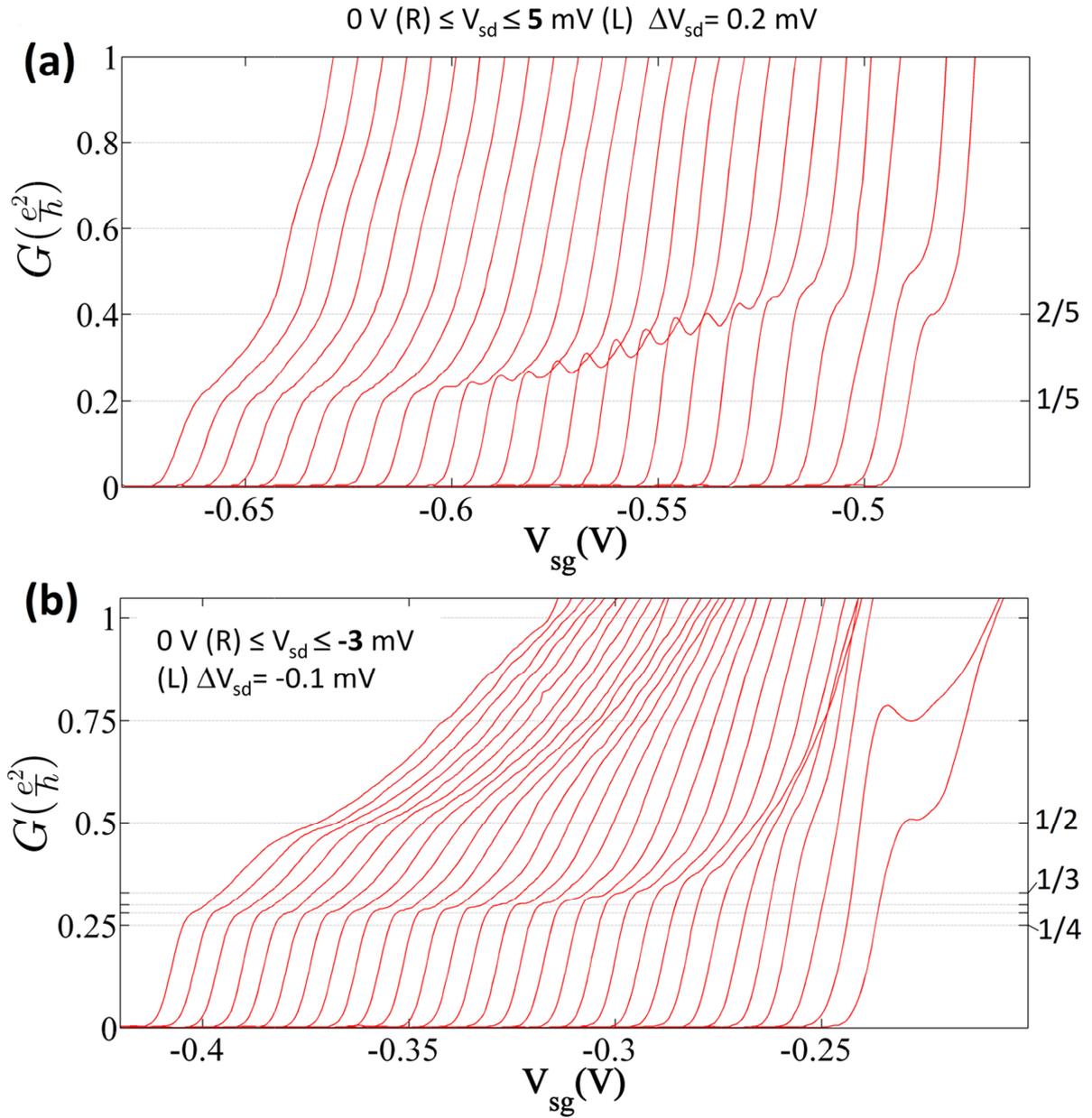

*Figure 4: Effect of dc source-drain bias on the fractional conductance. (a) The source-drain bias, $V_{sd}$ was incremented from 0 (right, R) to 5 mV (left, L) in steps of 0.2 mV; $\Delta V_{sg}$ =955 mV, $V_{tg}$= -0.36 V (sample 1). (b) Investigation of an Ohmic 1/2 structure; $V_{sd}$ was incremented from 0 (R) to -3 mV (L) in steps of -0.1 mV; $\Delta V_{sg}$ =0 mV, $V_{tg}$=-0.62 V (sample 2).*

The momentum of the electron system should remain constant for two conductance values where there is only a small change in confinement and is given by $N(E)m^*vdE$, where $N(E)$ is the density of states at the Fermi level, $v$ is the electron group velocity, $m^*$ is the electron effective mass and $dE=eV_{sd}$ is the energy spread of the injected electrons. For a small change in confinement potential and constant $V_{sd}$ the momentum should be constant, consequently, the difference in the current, $N(E)vdEe^*$, indicates a change in the effective charge $e^*$. However such confirmation requires a direct charge measurement as has been accomplished in measurements of shot noise [28].

There has been discussion of spin phases due to various patterns of ring exchange in the zig-zag [17, 18], and in 2D it has been proposed as a mechanism of fractionalization [29,30], although in the absence of a quantizing magnetic field the ring exchange can increase the electron energy [31]. However, in the quasi-1D zig-zag regime, in which a line of electrons splits into two, the transverse localization of the electrons contributes to the energy of the system and this can be reduced by the increase in spatial freedom given by a cyclic rotation. We suggest that this drives the correlated rotational motion responsible for the fractionalization. The various fractions observed correspond to different ring configurations which are dependent on the shape of the confinement potential and carrier concentration, these form a band separated by a gap from the main 1D band.

The pronounced and continuous changes in conductance with $B_{\parallel}$, from a small value, indicate that the system is responding to the change in flux rather than a simple spin polarization. With increasing field, we have identified minimum gradient plateaux at 1/6, 1/12, 1/18. This pattern may indicate a more complex pattern of cyclic motion involving pairing [32]. Although the effects of the interaction are most apparent in the plane of the electron gas there will be a correlated motion normal to the plane, to reduce the repulsion, which couples to the parallel magnetic field. In this context, it is interesting that studies of the 5/2 fraction in the FQHE have indicated that there is a dependence on the wavefunction normal to the plane [25].

Given the low carrier concentration used here, it is possible that the termination of the even-denominator series is due to all the electrons in the channel forming a collective state. The final rapid drop in the plateau values may indicate a change in mechanism to a dependence on the square of fractional charge. Measurement of charge [28] of the states identified here will be of importance in this regard. An interesting feature is the difference between these results and those where charge enters as a square as with holes in Ge [10], perhaps indicating a variation in the strength of the coupling due to the difference between s and p basis wavefunctions.

To summarize, at very low values of carrier concentration, <$6\times10^{10}$ cm$^{-2}$, we propose that the interactions between the electrons create correlated motion and fractional behavior in the absence of a magnetic field which is very different to that observed in the FQHE. There may be implications for quantum information schemes in the possible Non-Abelian nature of the 1/4 state and perhaps other even fractions. We have used the simplest structures for this investigation and the nature of the confinement may be altered by the electrons as with edge state reconstruction in the 2D case [33]. More sophisticated structures will allow a more precise probe of these new states and open new avenues in the design and tailoring of fractions with specific properties.

**Acknowledgements**
The work is funded by the Engineering and Physical Sciences Research Council, United Kingdom.

*sanjeev.kumar@ucl.ac.uk

concentration. Termination of the plateau is observed when the system moves to a new configuration with further change in confinement conditions.

## Supplementary Material

### 1. Sample details

The devices were fabricated on a GaAs/AlGaAs heterostructure grown using molecular beam epitaxy. The layer structure for sample 1 and sample 2 starting from the top, cap layer is given below:

Sample 1: GaAs cap(10 nm)/Al$_{0.33}$Ga$_{0.67}$As(200 nm)/GaAs(0.56 nm)/As:Si(8x10$^{11}$cm$^{-2}$)/ GaAs(0.56 nm)/Al$_{0.33}$Ga$_{0.67}$As (75 nm)/GaAs(buffer)(1000 nm).

Sample 2: GaAs cap(10 nm)/Al$_{0.33}$Ga$_{0.67}$As (doping 8x10$^{16}$ cm$^{-2}$) (200 nm)/Al$_{0.33}$Ga$_{0.67}$As (75 nm)/GaAs (buffer)(1000 nm).

The two-dimensional electron gas (2DEG) formed upto 290 nm beneath the surface had a low temperature mobility of $1.6 \times 10^6$ cm$^2$/V.s ($2.0 \times 10^6$ cm$^2$/V.s), and an electron density of $9.0 \times 10^{10}$ cm$^{-2}$ ($1.0 \times 10^{11}$ cm$^{-2}$) for sample 1 (sample 2). A pair of split gates of length (L) of 0.4 $\mu$m and width (W) of 0.7 $\mu$m, and a top gate of length 1 $\mu$m separated by a 300 nm thick insulating layer of cross-linked poly(methyl methacrylate) (PMMA) were patterned by a standard lithographic technique as shown in Fig. 1(s) below.

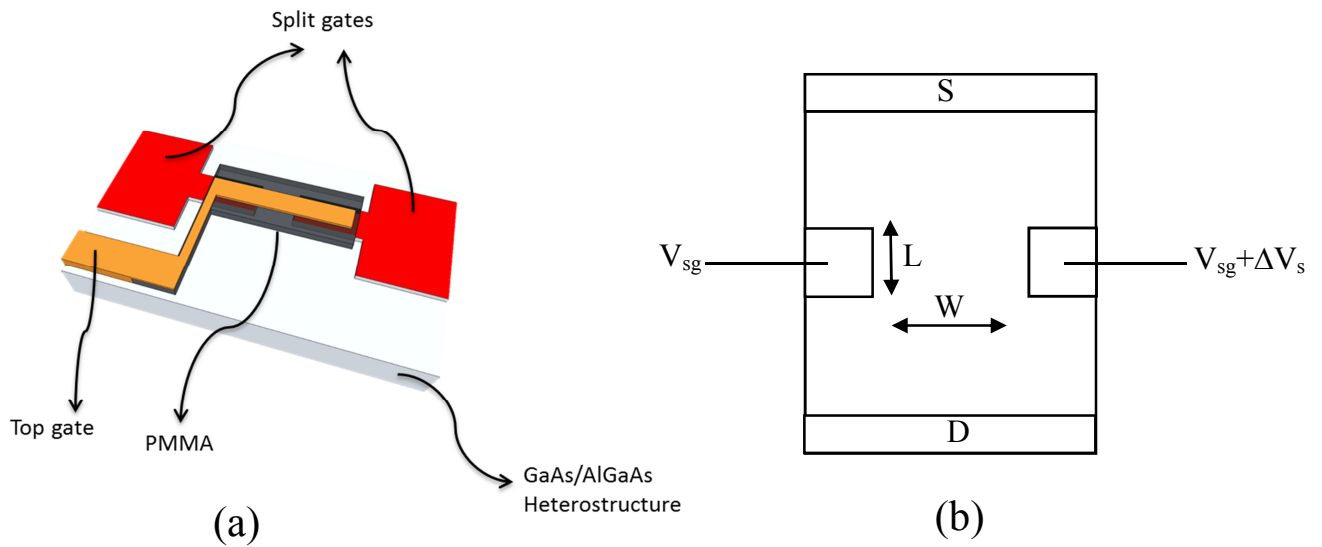

FIG. 1(s): Schematic diagram of the device. (a) The split gates are shown in red, PMMA is shown in grey, and the top gate in orange. The gates were deposited on GaAs/AlGaAs heterostructure. (b) Illustration of the experimental setup.

Defining a one-dimensional (1D) constriction in a 2DEG using split gates technique is well known. When a negative voltage $V_{sg}$ is applied to the split gates deposited on the top of a GaAs/Al$_x$Ga$_{1-x}$As heterostructure, the 2DEG underneath the gates is depleted resulting in a conducting 1D channel or constriction in the region between the split gates. When the elastic mean free path is more than the dimension (width (W) or length (L)) of the constriction, transport through the 1D constriction is ballistic and the differential conductance measured between source (S) and drain (D) via the two terminal method is given by, G= N(2e$^2$/h), where N=1,2,3…represents 1D subbands. This is the case of strong confinement where electrons are non-interacting. The confinement is weakened by applying a negative voltage on the top gate, and then repeating the split gates voltage sweeps to measure the conductance in a weaker confinement. The two-terminal differential conductance (*G*) measurements were performed in the presence of an excitation voltage of 10 $\mu$V at 73 Hz in a cryofree dilution refrigerator with a lattice temperature of 25 mK.

The measurement with asymmetry biasing on the split gates is shown in Fig. 1, main text, where conductance on a 1D channel is measured as a function of increasing the asymmetry of the confinement potential by applying an offset, $\Delta V_{sg}$, to one of the split gates, [for setup, see Fig. 1(s)], with the two gates then being swept together. The net effect of the offset was to push the conducting channel sideways and increase its width.

## 2.     **The 1/2 and 1/4 states**

Figure 3 in the main text illustrates a portion of the sensitivity of the system to a change in carrier concentration at a fixed in-plane magnetic field of 10 T. The in-plane magnetic field was normal to the current direction. Here we show a detailed result. There is a constant offset applied on the split gates ($\Delta V_{sg}$= -180 mV) producing the asymmetry of confinement potential; the top gate produces a decreasing carrier concentration in the 1D channel, the effect is shown from left ($V_{tg}$=-0.39 V) to right ($V_{tg}$=-0.62 V) in Fig 2(s). Due to the presence of a large magnetic field of 10 T, the usual spin polarized plateau at 1 is seen throughout the confinement spectrum in Fig. 2(s). On reducing the carrier concentration, a structure close to 2/3 appears which strengthens and a flat plateau develops at 1/2. The traces on either side of the indexed 1/2 exhibit flat plateau in the range 6/11 to 5/11, which indicates a band of stable states within these limits. On further lowering the carrier concentration, the structure drops in value and flattens at 1/4 before disappearing indicating a transition to 2D behavior. Flat plateaux form around the 1/4 between limits of 3/10 and 6/25 indicating that there is a band of stable states in a similar way to the states around the 1/2. The similarity in the behavior of the structure near the 1/2 and 1/4 values found here indicates that they are linked. The flattening of the plateaux around 1/2 and 1/4 gives us confidence that the effect is free from impurities.

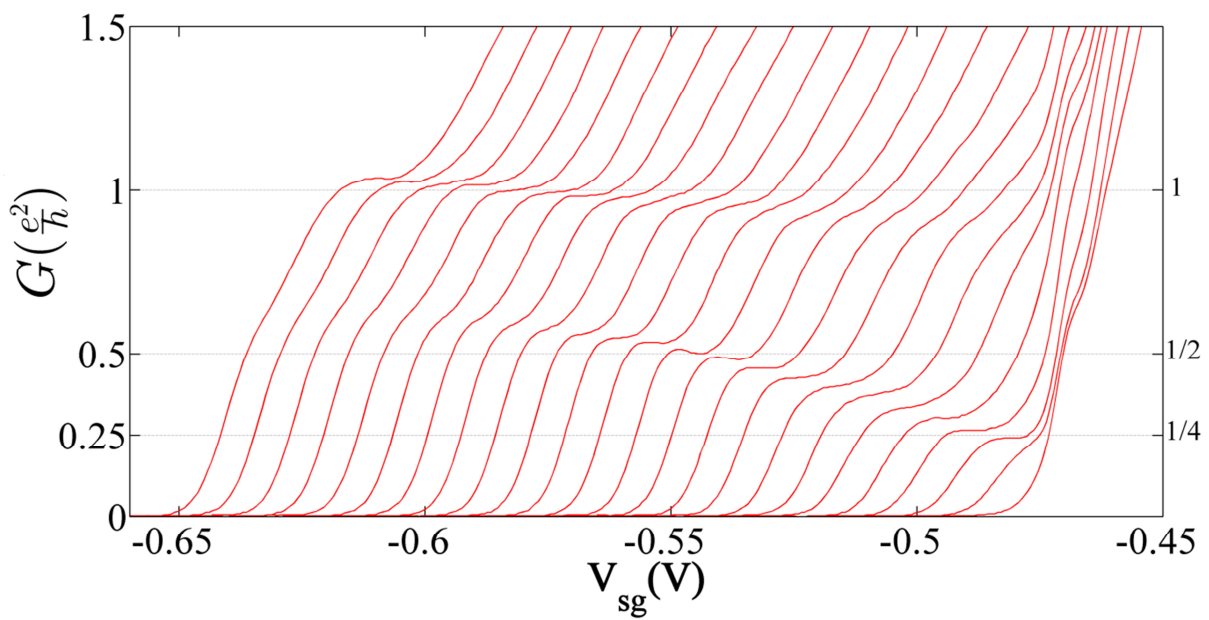

FIG. 2(s): Effect of variation of carrier concentration on the conductance in the presence of $B_{||}$ (sample-2). $V_{tg}$ was varied from -0.39 V (left) to -0.62 V (right) in steps of -10 mV, at $B_{||}$=10 T; $\Delta V_{sg}$ =-180 mV.

### 3. Source-drain bias

Figure 4(b) in the manuscript shows the differential conductance of an initially Ohmic fractional plateau at 1/2, Fig. 3(s), below, is an enhanced view of part of Fig. 4(b). The value of differential conductance drops, as expected with increasing $V_{sd}$ but forms a plateau at 1/3 followed by a flattening at 3/10 with intervening structure. Finally before disappearing the structure attempts to converge at 2/7 remaining above the value of 1/4 produced by the lifting of the momentum degeneracy.

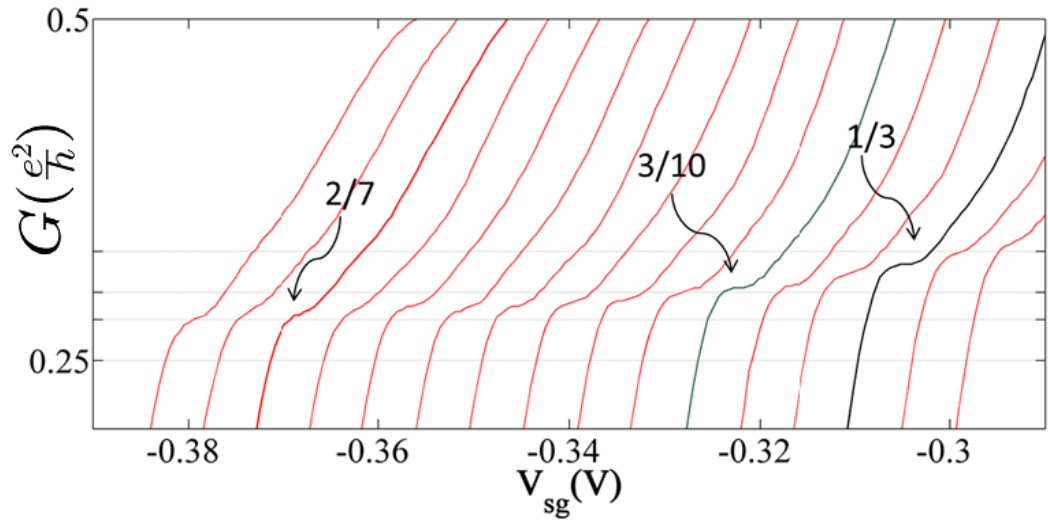

FIG. 3(s): Effect of dc source-drain bias on the fractional state. A zoom-out of the data in Fig. 4(b) in the manuscript is shown here (sample-2).